\documentclass{jfm}
\usepackage[utf8]{inputenc}

\renewcommand{\pagelimitfooter}{}
\renewcommand{\absfooterflag}{}

\newcounter{ncorresp}
\let\Oldcorresp\corresp
\renewcommand{\corresp}[1]{{\Oldcorresp{#1}}\stepcounter{ncorresp}}

\usepackage{aas_macros}
\usepackage{amsmath}
\usepackage{amssymb}
\usepackage{mathtools} 

\usepackage{graphicx}
\usepackage{newtxtext}
\usepackage{newtxmath}
\usepackage{natbib}
\usepackage{hyperref}
\hypersetup{
    colorlinks = true,
    urlcolor   = blue,
    citecolor  = black,
}

\newcommand{\RomanNumeralCaps}[1]
\linenumbers

\renewcommand{\nu}{\mu}

\renewcommand{\vec}[1]{\boldsymbol{#1}}

\usepackage[inline]{enumitem} 
\newenvironment{enuminline}{\begin{enumerate*}[label=(\roman*)]}{\end{enumerate*}} 

\usepackage{siunitx} 

\renewcommand{\dh}{\partial}
\renewcommand{\d}{\mathrm{d}}

\newcommand{\dive}{\nabla\!\cdot}

\newcommand{\grad}{\nabla}

\newcommand{\defn}{\equiv}
\newcommand{\abs}[1]{\left|#1\right|}

\newcommand{\scalesAs}{\sim}
\newcommand{\orderOf}{\sim}

\newcommand{\shBg}[1]{{#1}_0} 
\newcommand{\shFl}[1]{{#1}'} 
\newcommand{\shFt}[2]{{#1}_{#2}} 
\renewcommand{\sh}[3]{{#1}_{#2, #3}} 

\newcommand{\shdZ}{\Delta Z_m} 

\newcommand{\nlterm}{level coupling}

\newcommand{\R}{\Rey} 
\renewcommand{\Pr}{\Pran} 
\newcommand{\Gr}{\mbox{\textit{Gr}}} 
\newcommand{\Ra}{\mbox{\textit{Ra}}} 
\newcommand{\Pe}{\Pen} 

\title{Shell model for stratified convection: implications for the solar convective conundrum}
\author{
Kishore Gopalakrishnan
\corresp{\email{kishoreg@iucaa.in}}
\and
Nishant K. Singh
\corresp{\email{nishant@iucaa.in}}
}
\affiliation{Inter-University Centre for Astronomy \& Astrophysics, Post Bag 4, Ganeshkhind, Pune 411 007, India}

\begin{document}
\maketitle
\addtocounter{footnote}{\value{ncorresp}}

\begin{abstract}
	We extend the notion of a shell model to stratified systems, and propose one that represents stratified, nonmagnetic, nonrotating convection at low Mach number.
	Motivated by profiles of background stratification that support convection in stars such as the Sun, we study numerical solutions corresponding to a highly unstable layer above a mildly unstable layer.
	We find that at low Prandtl number, convective amplitudes decrease with depth in the lower layer.
	This suggests that the suppression of convection in the deeper layers of the Sun's convection zone (the convective conundrum) can be addressed without necessarily appealing to rotation or magnetic fields.
\end{abstract}



\section{Introduction}

Turbulent convection in stars is often described by the mixing length theory \citep[chapter 6]{bohmvitenseBookVol3}.
In this theory, convection is restricted to unstably stratified regions of a star, and energy is predominantly transported by eddies of the order of the local scale height.
The entirety of the Sun's convection zone is unstably stratified (\citealt{SolarModelS}; \citealt{SchumacherSreenivasanSolConv2020}, fig.~3), and thus one expects solar convection to be dominated by length scales of the order of the scale height deep in the convection zone.
This is also supported by simulations of deep solar convection \citep[e.g.\@][]{miesch2008ASH}.

However, recent helioseismic analyses \citep{hanasoge2010, hanasoge2012anomalously, hanasoge2020} suggest that the kinetic energy at large scales deep in the Sun's convection zone is much lower than expected.\footnote{
There is some controversy around these observations; for reviews, see \cite{hanasoge2016}; and p.~40 of \cite{christensen2021solar}.
}
Such suppression also seems to resolve other issues:
\citet{Lord2014} find that the observed photospheric spectra at supergranular and larger scales can be explained if convection in deeper layers is suppressed at large scales;
further, suppression of convective velocities deep in the convection zone would allow simulations to reproduce solar differential rotation, where the equator rotates faster than higher latitudes \citep{kapyla2014bistableDiffRot, guerrero2013, gastine2013}.
The mismatch between observations and simulations is often dubbed the `convective conundrum';
a variety of explanations have been proposed for it.

According to one proposal, the deep convection zone is dominated by cool downflowing plumes  (`entropy rain') from a highly unstable layer near the surface \citep{Spruit1997, brandenburg2016stellar, Cossette2016}.
In this picture, the length scale of convection is set by the scale height at the surface or by the depth of the top unstable layer, and so convective flows need not be excited at large scales deep in the convection zone.
Possible reasons for not observing such downflows in simulations of deep convection are that
\begin{enuminline}
	\item enhanced diffusivities suppress these small-scale flows; and
	\item near-surface layers are typically not included in such simulations.
\end{enuminline}
However, \citet{hotta2019weakSurface} claim that the mere inclusion of near-surface layers in simulations is not sufficient to explain the convective conundrum.

Another proposed solution to the convective conundrum is that the flows are suppressed by small-scale magnetic fields \citep{Fan2014, Hotta2015SSD, hotta2021highRes}.
The small-scale magnetic fields supposedly result in an enhanced turbulent Prandtl number (the Prandtl number, \Pr{}, is the ratio of the kinematic viscosity to the thermal diffusivity; the latter may include both conduction and radiation), which may then suppress the large-scale velocities \citep{omara2016turbPr, karak2018highPrt}.
However, it is still unclear \citep{karak2018highPrt} if an enhanced turbulent Prandtl number can reproduce the differential rotation profile observed in the Sun.

A third proposal is that rotation suppresses convective flows at the largest scales \citep{Featherstone2016}.
However, arguments about the influence of rotation typically make assumptions about the convective amplitudes, and thus it is not clear if rotation alone is enough to suppress the convective velocities, or if the velocities get suppressed by some other effect, allowing rotation to become important.

More promisingly, \citet{Featherstone2016HighRa} report that the kinetic energy at large scales decreases as the Rayleigh number (\Ra{}, a measure of how strongly convection is driven) increases.
Since $\Ra{} \orderOf \num{e20}$ in the Sun \citep{SchumacherSreenivasanSolConv2020}, this
may explain, to some extent, the convective conundrum.
However, \citet{kapyla2021Pr} find that the spectral distribution of velocity is insensitive to \Ra{}.
This discrepancy might be due to the different sub-grid-scale (SGS) models they use. 
The SGS diffusion in \citet{Featherstone2016HighRa} is applied to the total entropy, whereas \citet{kapyla2021Pr} applies it only to the entropy fluctuations such that it does not contribute
to the mean energy flux.

Geophysical and laboratory flows are typically at Prandtl numbers (\Pr{}) close to or much larger than one.
On the other hand, stellar convection takes place at very low \Pr{}; e.g. in the solar convection zone, $\Pr \orderOf \num{e-6}$ \citep{SchumacherSreenivasanSolConv2020}.
It is difficult to run simulations with \Pr{} significantly different from unity, due to the wide range of scales that need to be simultaneously resolved.
Moreover, the widely-used anelastic approximation exhibits spurious instabilities in rotating systems at low \Pr{} \citep{calkins2015anelastic}.
Most 3D simulations of stellar convection are run with enhanced diffusivities and $\Pr\orderOf 1$; e.g.\@ the ASH simulations \citep{miesch2008ASH} use $\Pr=0.25$.
Others \citep{hotta2021highRes, hotta2019weakSurface} use artificial diffusion schemes, but still seem to have $\Pr\orderOf 1$.

Considering $0.1 \le \Pr \le 10$,
\citet{kapyla2021Pr} finds that despite the velocity spectra being rather insensitive to \Pr{}, the
nature of convection changes significantly, with stronger
downflows and larger overshoot depths at low \Pr{}.
Convection at low \Pr{} appears very turbulent, yet less efficient: with decreasing \Pr{}, larger velocity amplitudes are needed to convect the same amount of energy.

It seems to be common to make assertions about the nature of low-\Pr{} convection based on the simplified system of equations derived by \citet{spiegel1962lowPr}.
However, we note (as recognized by \citet{lignieres99}) that the aforementioned equations are only valid at low Peclet number (\Pe{}, the ratio of the thermal diffusion timescale to the advection timescale);
they are thus not applicable to the solar convection zone, where $\Pe = \R\Pr \orderOf \num{e7}$ \citep{SchumacherSreenivasanSolConv2020}.
We are not aware of any simplified versions of the equations governing convection in the simultaneous limit of high Peclet number and low Prandtl number.

Given the complexity of solar convection and the infeasibility of fully resolved simulations, simplified models of convection help us understand which effects are important.
We present one such simplified model, based on a generalization of `shell models'.
Shell models \citep{yamada1987shell, brandenburg1992shell, lvov1998shell} are simplified models of turbulence that have proved successful in deepening our understanding of energy cascades in homogeneous, isotropic turbulence (see \citealt{biferale2003shell} for a review).
Shell models representing convection in a thin layer (the Boussinesq approximation) have been studied in the past \citep{brandenburg1992shell, kumar15shell}.
While shell models are themselves quite interesting, our attitude towards them is that they are relatively simple systems that nevertheless qualitatively capture some aspects of turbulent energy cascades.

In this paper, we study a generalized shell model,
where the shell amplitudes are functions of both the depth and the wavenumber,
without considering rotation and magnetic fields. 
In a setup consisting of a highly unstable layer above a weakly unstable layer, we find that simply using a low enough Prandtl number causes the velocity amplitudes to decrease with depth.
Our results suggest that modelling the effect of a low Prandtl number is likely to be essential for a complete explanation of the convective conundrum.

In section \ref{section: formulation}, we explain how we extend the notion of a shell model to a stratified system.
In section \ref{section: results}, we present numerical solutions of such a model and discuss their implications for solar convection.
Finally, in section \ref{section: conclusions}, we state our main conclusions and suggest future avenues of research.

\section{Formulation of a shell model for stratified convection}
\label{section: formulation}

\subsection{Simplifying the equations of motion}
Let us start with the following equations of motion for an ideal gas in a nonrotating plane-parallel domain without magnetic fields:
\begin{align}
	\begin{split}
		\frac{\dh \rho }{\dh t} 
		={}&
		- \dive{\left(  \rho \vec{v} \right)}
	\end{split}
	\\
	\begin{split}
		\frac{\dh \vec{v} }{\dh t} 
		={}& 
		- \left(\vec{v}\cdot\nabla\right)\!\vec{v}
		- \frac{ \nabla p }{\rho}
		+ \nu \, \nabla^2 \vec{v}
		+ \vec{g}
	\end{split}
	\\
	\begin{split}
		\frac{\dh s }{\dh t} 
		={}&
		- \vec{v} \cdot\nabla s
		+ \frac{\kappa C_P}{T} \, \nabla^2 T
	\end{split}
	\\
	\begin{split}
		T
		\propto{}& 
		\rho^{\gamma-1} \exp{\left( \frac{s}{C_V} \right)}
	\end{split}
	\\
	\begin{split}
		p ={}& 
		\rho R T
	\end{split}
\end{align}
where $\rho$ is the density;
$\vec{v}$ is the velocity;
$p$ is the pressure;
$\nu$ is the kinematic viscosity;
$\vec{g}$ is the acceleration due to gravity;
$s$ is the specific entropy;
$\kappa$ is the thermal diffusivity (which may include both conduction and radiation);
$T$ is the temperature;
$C_P$ is the specific heat at constant pressure;
$C_V$ is the specific heat at constant volume;
$\gamma \defn C_P/C_V$;
$R \defn C_P - C_V$;
and $t$ is the time.
We have neglected the effects of compressibility on viscous dissipation; neglected viscous heating; neglected the effect of density variations on the thermal conductivity; neglected internal heat sources/sinks; and assumed $\nu$, $\kappa$, $R$, and $C_P$ are constant throughout the domain.

Now, for every quantity $\Box$ ($= \rho, \vec{v}, s, T, p$), we perform the split,
\begin{equation}
	\Box = \shBg{\Box} + \shFl{\Box}\,,
\end{equation}
where $\shBg{\Box}$ represents a `background', and $\shFl{\Box}$ represents deviations from this background.
We assume the problem contains two disparate spatial scales, such that $\shFl{\Box}$ can be treated as a function of both a `small-scale' coordinate and a `large-scale' vertical coordinate, while $\shBg{\Box}$ is only a function of the `large-scale' coordinate.
Our basic approach will be to Fourier-transform the `small-scale coordinate' and bin it into shells, so that we end up with a shell model whose dynamical variables depend both on the shell index and on a `large-scale' coordinate that represents the stratification.

We assume $\shBg{\vec{v}} = 0$, so that $\shFl{\vec{v}} = \vec{v}$.
Further, we assume the background quantities are prescribed solutions of the equations of motion, and do not evolve with time (i.e. the background state is in hydrostatic equilibrium, and satisfies the ideal gas equation).
Dropping various terms which we expect to be nonessential,\footnote{
This is of course highly subjective and difficult to justify a priori; however, we will see in section \ref{section: results} that we do manage to reproduce the effect we are interested in.
} recalling that the background is in hydrostatic equilibrium, and simplifying some of the more complicated terms, we write the equations of motion as
\begin{align}
	\begin{split}
		\frac{\dh \shFl{\rho} }{\dh t} 
		={}&
		- \dive{\left(  \shBg{\rho} \vec{v} \right)}
		- \dive{\left(  \shFl{\rho} \vec{v} \right)}
	\end{split}
	\\
	\begin{split}
		\frac{\dh \vec{v} }{\dh t} 
		={}& 
		- \left(\vec{v}\cdot\nabla\right)\!\vec{v}
		- \frac{ \nabla \shFl{p} }{\shBg{\rho} }
		+ \nu \, \nabla^2 \vec{v}
		+ \frac{ \shFl{\rho} \vec{g} }{\shBg{\rho} }
	\end{split}
	\\
	\begin{split}
		\frac{\dh \shFl{s} }{\dh t} 
		={}&
		- \vec{v} \cdot\nabla \shBg{s}
		- \vec{v} \cdot\nabla \shFl{s}
		+ \frac{\kappa C_P}{ \shBg{T} } \, \nabla^2 \shFl{T} 
	\end{split}
	\\
	\begin{split}
		\shFl{T} 
		={}& 
		\shBg{T} \left[ \left( \gamma - 1\right) \frac{\shFl{\rho} }{\shBg{\rho} } + \frac{\shFl{s} }{C_V} \right]
	\end{split}
	\\
	\begin{split}
		\shFl{p} ={}& 
		\shFl{\rho} R \shBg{T} + \shBg{\rho} R \shFl{T} 
	\end{split}
\end{align}

The procedure we will describe in section \ref{section: constructing shell model} can be carried out for the above set of equations; however, they turn out to be numerically unstable unless a dissipative term is added to the continuity equation (for an example of the usage of such terms, see \citet{Gent2021superSSD}). 
We avoid this as the addition of such a term will complicate the interpretation of the resulting shell model;
in fact, it is suspected \citep[p.~2]{brandenburg2016stellar} that such artificially enhanced diffusivities are the reason numerical simulations of the deep solar convection zone don't agree with helioseismic observations.
Instead, we work in the limit of infinite sound speed, i.e.\@ we neglect the pressure fluctuations, which immediately leads to
\begin{equation}
	\frac{\shFl{\rho} }{\shBg{\rho} } \approx - \frac{\shFl{T} }{\shBg{T} }
\end{equation}
One may thus consider the system of equations
\begin{align}
	\begin{split}
		\frac{\dh \vec{v} }{\dh t} 
		={}& 
		- \left(\vec{v}\cdot\nabla\right)\!\vec{v}
		+ \nu \, \nabla^2 \vec{v}
		- \frac{ \shFl{s} \vec{g} }{C_P}
	\end{split}
	\\
	\begin{split}
		\frac{\dh \shFl{s} }{\dh t} 
		={}&
		- \vec{v} \cdot\nabla \shBg{s}
		- \vec{v} \cdot\nabla \shFl{s}
		+ \frac{\kappa}{ \shBg{T} } \, \nabla^2{\left( \shBg{T} \shFl{s}  \right)}
	\end{split}
\end{align}
It turns out (on examining numerical solutions of the resulting shell model, not presented here) that the behaviour of this system (at least the aspects we are interested in) is not qualitatively changed by neglecting the variation of $\shBg{T}$ in the entropy equation, so in what follows, we work with the system
\begin{align}
	\begin{split}
		\frac{\dh \vec{v} }{\dh t} 
		={}& 
		- \left(\vec{v}\cdot\nabla\right)\!\vec{v}
		+ \nu \, \nabla^2 \vec{v}
		- \frac{ \shFl{s} \vec{g} }{C_P}
	\end{split}
	\\
	\begin{split}
		\frac{\dh \shFl{s} }{\dh t} 
		={}&
		- \vec{v} \cdot\nabla \shBg{s}
		- \vec{v} \cdot\nabla \shFl{s}
		+ \kappa \, \nabla^2 \shFl{s}
	\end{split}
\end{align}

\subsection{Constructing the shell model}
\label{section: constructing shell model}
We now Fourier-transform the small-scale variable (say $\vec{r}$) to the wavevector $\vec{k}$, and restrict the allowed values of $k \defn \abs{\vec{k}}$ to
\begin{equation}
	k_n = k_1 h^{n-1} \quad\text{, where } n = 1,2,3,\dots, N\,,
\end{equation}
where $h$ is called the inter-shell ratio.
We further assume all $\vec{k}$-dependent quantities depend only on $k$, and replace convolutions over the small-scale wavevector by summations over neighbouring shells.
While replacing convolutions, the choice of which shells to sum over should be consistent with the relevant conservation laws.
The Fourier transform of every quantity $\shFl{\Box}(\vec{r},Z)$ is now denoted by $\shFt{\Box}{n}(Z)$ (where $Z$ is the large-scale coordinate).
We use $\shBg{\Box}(Z)$ to denote the (prescribed) background quantities, as before.
Now, we replace the continuous variable $Z$ by a set of discrete `levels' which are indexed by $m=1,2,3,\dots,M$.
For all variables $\Box$, we take $\shFt{\Box}{n}(Z) \to \sh{\Box}{n}{m}$.
Derivatives wrt.\@ $Z$ are replaced by second-order finite differences, so that, e.g.\@
\begin{equation}
	\frac{\d \shBg{s} }{\d Z} \to \frac{ \sh{s}{0}{m+1} - \sh{s}{0}{m-1} }{2 \shdZ}
\end{equation}
where $\shdZ$ is a `level spacing', which we assume to be a constant for the sake of simplicity.
To specify the boundary conditions, we add `ghost levels' for which $m=0,M+1$.
On these ghost levels, all the fluctuating quantities are set to zero, while the background quantities are set such that their gradient at the boundary is constant.
For example,
\begin{equation}
	\sh{s}{n}{0} =
	\begin{dcases}
		0 &\, , n\ne 0 \\
		2 \sh{s}{n}{1} - \sh{s}{n}{2} & \, , n=0
	\end{dcases}
\end{equation}

We then end up with the equations
\begin{subequations}
\begin{align}
	\begin{split}
		\frac{\dh \sh{v}{n}{m} }{\dh t} 
		={}& 
		i C_1 \, k_{n+1} \sh{v}{n+2}{m} \sh{v}{n+1}{m}^*
		+ i C_2 \, k_{n} \sh{v}{n+1}{m} \sh{v}{n-1}{m}^*
		- i C_3 \, k_{n-1} \sh{v}{n-1}{m} \sh{v}{n-2}{m}
		\\& - \nu k_n^2 \sh{v}{n}{m}
		- \frac{ \sh{s}{n}{m} g }{ C_P }
	\end{split}
	\\
	\begin{split}
		\frac{\dh \sh{s}{n}{m} }{\dh t} 
		={}&
		i C_1 \, k_{n+1} \sh{v}{n+1}{m}^* \sh{s}{n+2}{m}
		+ i C_2 \, k_{n} \sh{v}{n-1}{m}^* \sh{s}{n+1}{m}
		- i C_3 \, k_{n-1} \sh{v}{n-1}{m} \sh{s}{n-2}{m}
		\\& - \kappa k_n^2 \sh{s}{n}{m}
		+ \underbrace{ \kappa \, \frac{ \sh{s}{n}{m+1} - 2 \sh{s}{n}{m} + \sh{s}{n}{m-1} }{\shdZ^2 } }_\text{\nlterm} {}
		- \sh{v}{n}{m} \, \frac{ \sh{s}{0}{m+1} - \sh{s}{0}{m-1} }{2 \shdZ}
	\end{split} \label{eq: shell evolution equation entropy}
\end{align} \label{eq: shell evolution equations}%
\end{subequations}
The terms linking different shells above have been chosen such that the linkages in the momentum equation are the same as those in the `Sabra' \citep{lvov1998shell} model, while the linkages in the entropy equation have been chosen to retain the conservation laws for $\rho \vec{v}$ and $\rho s$ (before taking the limit of infinite sound speed).
Note that coupling between different vertical levels is only introduced by the term marked `\nlterm{}' above.

To solve equations \ref{eq: shell evolution equations}a--b, we need to specify $N$, $M$, $h$, $C_1$, $C_2$, $C_3$, $\nu$, $\gamma$, $g$, and $\shdZ$.
The background is completely specified by the arbitrary choice of $\sh{s}{0}{m}$.
Initial values of $\sh{s}{n}{m}$ and $\sh{v}{n}{m}$ need to be specified.
Henceforth, we set $C_1 = 1$, $C_2 = -0.5$, $C_3 = -0.5$, and $h=2$;
this choice of coefficients was used to represent 3D homogeneous isotropic turbulence by \citet{lvov1998shell}.

In appendix \ref{appendix: Pr Ra in shell model}, we list formulae that allow us to estimate the Rayleigh, Prandtl, and Reynolds numbers given the values of model parameters.

\section{Suppression of deep convection at low Prandtl number}
\label{section: results}

We now present some solutions of equations \ref{eq: shell evolution equations}.
To solve them, we use the BDF solver \citep{byrne1975polyalgorithm,shampine1997matlab} from SciPy \citep{scipy2020}; this is an implicit solver for stiff equations that can work in the complex domain.
We evolve the equations starting from a nonzero initial condition, and average the required quantities (e.g.\@ $\abs{v_n}^2$) over a time interval much larger than the correlation time of the shell variables.

\begin{figure}
	\centering
	\includegraphics[width=0.6\textwidth,keepaspectratio=true]{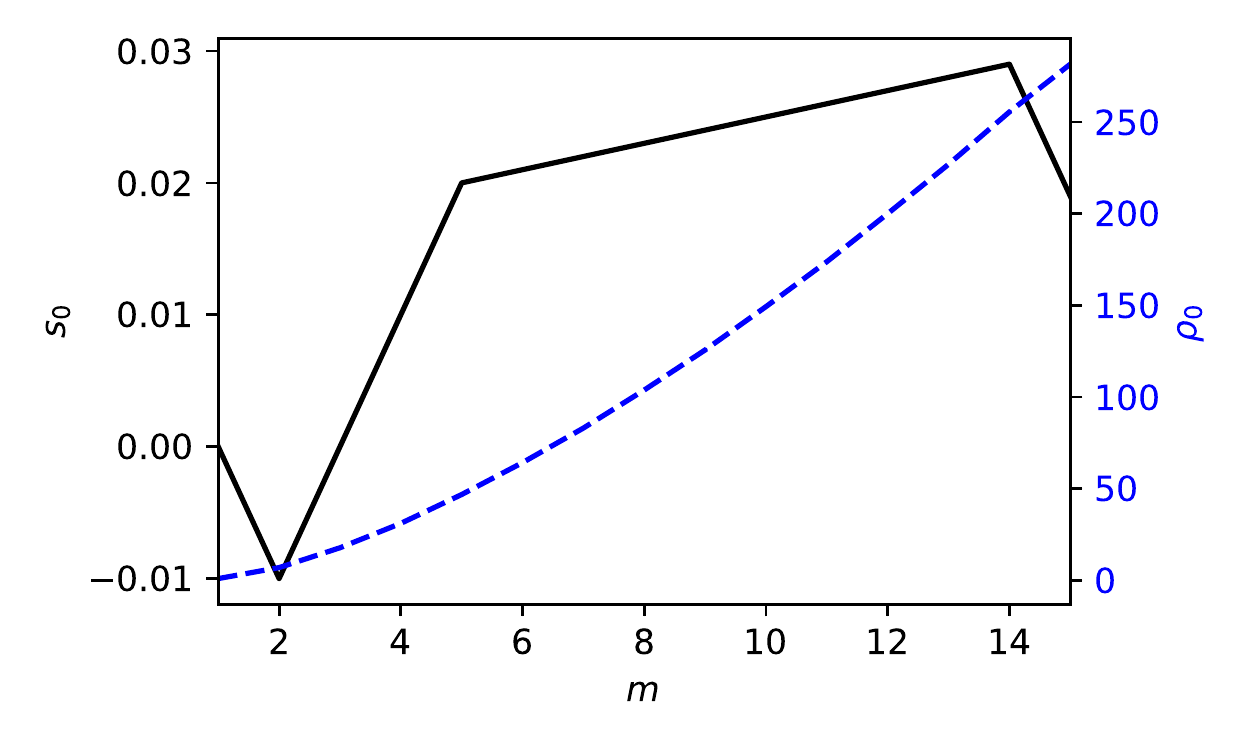}
	\caption{
	Depth dependence of the background profiles used for all the cases presented ($m=1$: top
	layer; $m=15$: bottom layer).
	Solid (black): entropy; dashed (blue): density.
	}
	\label{fig: background profile}
\end{figure}

For all the solutions presented here, we set the background entropy profile (figure \ref{fig: background profile}) to have a thin unstable layer with a large entropy gradient, on top of a thick unstable layer with a smaller entropy gradient; additional thin and stably stratified layers are placed at the top and bottom boundaries in order to maintain consistency with the boundary conditions.
Initially, we set $s=1$ for the first four shells at $m=3$, with the remaining dynamical variables being set to zero.
We set $N=30$; $M=15$; $\gamma=5/3$; $\shdZ=1$, $C_P=1$, $\kappa=1$.
As described in appendix \ref{appendix: Pr Ra in shell model}, the remaining parameters ($\nu$ and $g$) can be used to control \Ra{} and \Pr{}.
For the $\Pr = 1$ case, we set $g=\num{e6}$ and $\nu=1$.
For the other cases, $\nu$ and $g$ are varied as described in appendix \ref{appendix: vary Pr keeping Ra const}.

\begin{figure}
	\centering
	\includegraphics[width=\textwidth,keepaspectratio=true]{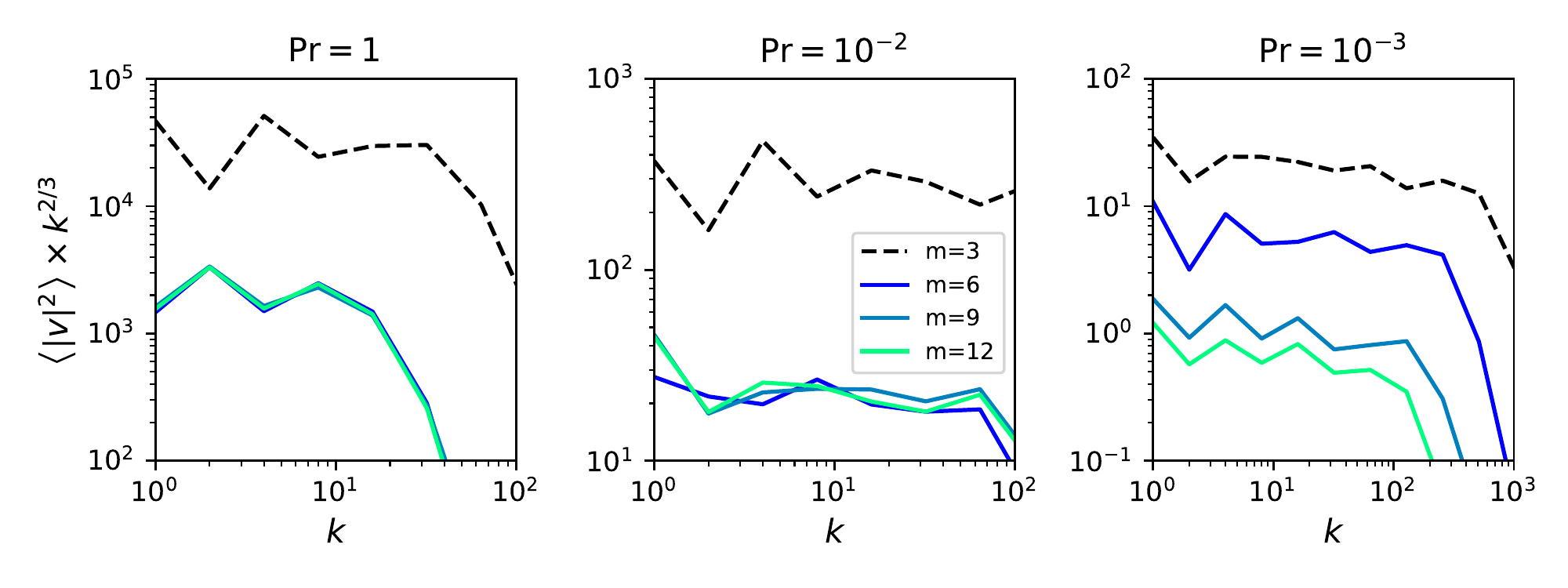}
	\caption{
	Spectra at various vertical levels for $\Ra = \num{8.6e7}$.
	Larger values of $m$ are deeper in the domain; \Pr{} decreases from left to right.
	Only relative magnitudes of the spectra at various depths are important here, as their absolute values are affected by the process of varying \Pr{} while keeping \Ra{} constant; see appendix \ref{appendix: vary Pr keeping Ra const}.
	}
	\label{fig: compare spectra different Pr}
\end{figure}
\begin{figure}
	\centering
	\includegraphics[width=\textwidth,keepaspectratio=true]{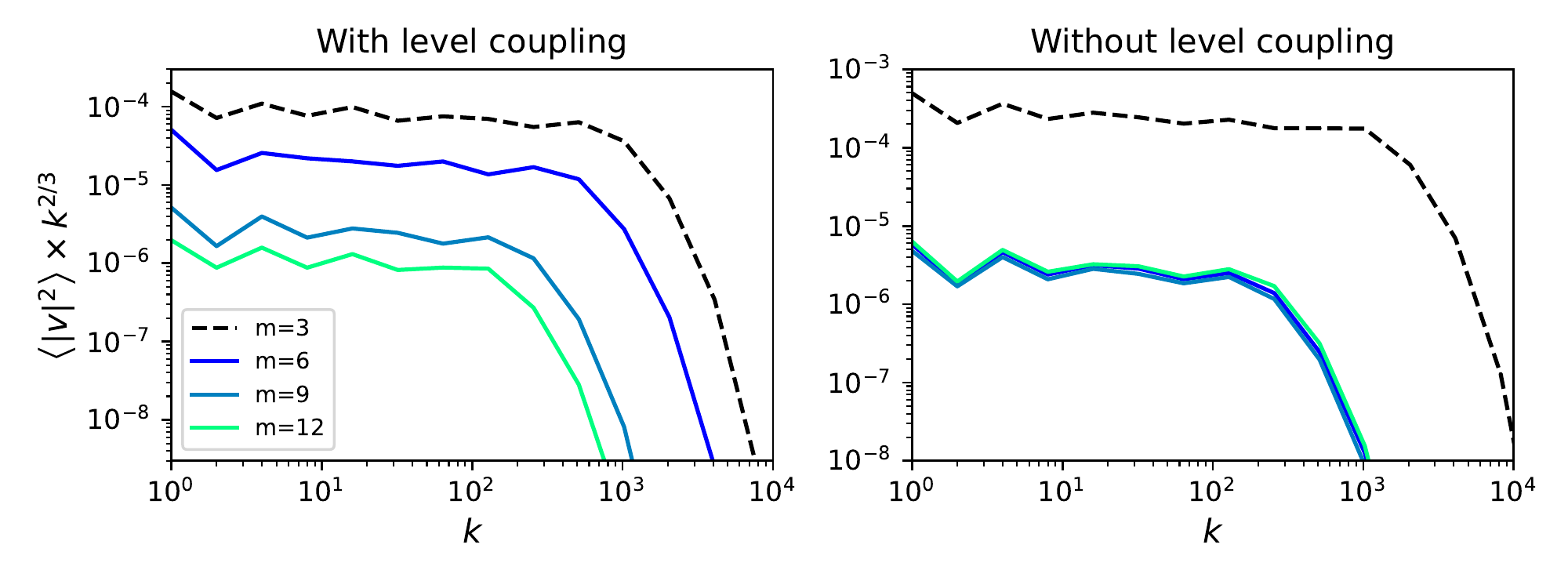}
	\caption{
	Effect of the \nlterm{} term (see equation \ref{eq: shell evolution equation entropy}) at $\Pr = \num{e-6}$ and $\Ra = \num{8.6e7}$.
	}
	\label{fig: spectrum without level coupling}
\end{figure}

First of all, we note that in an unstratified shell model (that represents homogeneous, isotropic, incompressible turbulence), one expects $\big< \abs{v_n}^2 \big> \scalesAs k_n^{-2/3}$ \citep[eq.~14]{biferale2003shell},\footnote{
Since the shells are logarithmically spaced, the KE spectrum is actually $\big< \abs{v_n}^2/k_n \big>$, which indeed scales as $k_n^{-5/3}$.
} where the angle bracket denotes a time average.
We find that the velocity spectra follow the same scaling with wavenumber that is expected in the unstratified case (e.g.\@ all three panels of figure \ref{fig: compare spectra different Pr}).
This is perhaps not too surprising, as we have intentionally kept the shell couplings similar to those in the unstratified case.
It is possible that a more careful consideration of symmetries and conservation laws suggests a different form of the shell coupling, which could lead to a different power-spectral slope.

In figure \ref{fig: compare spectra different Pr}, we see that simply decreasing \Pr{} (keeping \Ra{} fixed) is enough to cause convective amplitudes to decrease with depth.
What we observe in this model is that the spectra remain roughly Kolmogorov at all depths, but that their amplitudes depend on the depth at which they reside.

In figure \ref{fig: spectrum without level coupling}, we see that when the `\nlterm{}' term is dropped from equation \ref{eq: shell evolution equation entropy}, lowering \Pr{} does not result in suppression of large scales.
This suggests that along with low \Pr{}, nonlocality is also crucial to the effect we observe.

A comparison of the three panels in figure \ref{fig: compare spectra different Pr} suggests that the effect of low \Pr{} is to boost the spectra at intermediate depths, rather than to suppress convection in deeper layers.
However, note that in all these cases, the background entropy gradient is the same.
In simulations or real systems where the total energy transport needs to be constant, the effect observed here may manifest as an apparent suppression of the spectra with depth.

The behaviour described here (enhanced influence of surface layers at low \Pr{}) is not affected by whether the lower layers are stable or unstable; it is sufficient for the absolute value of the entropy gradient in the lower layers to be much less than that in the upper layers.

A caveat is in order:
note that all the spectra exhibit oscillations (in the wavenumber-domain) at small wavenumbers.
These oscillations are due to the flux of a conserved quantity with dimensions of kinetic helicity (\citealt{lvov1998shell}, p.~1813; \citealt{DitlevsenShellBook}, p.~101).
In forced shell models, one typically eliminates these oscillations by choosing a particular kind of forcing; however, we do not have the freedom to do that in our system.
Such (presumably spurious) features seem to be quite common in previous shell models for Boussinesq convection (\citealt{brandenburg1992shell}, fig.~1; \citealt{kumar15shell}, fig.~4)\footnote{
Interestingly, it seems that the introduction of a magnetic field gets rid of such oscillations \citep[fig.~2]{brandenburg1992shell}.
This may be related to the fact that the kinetic helicity is not conserved in the presence of magnetic fields.
}\footnote{
\citet{kumar15shell} seem to dismiss these oscillations as ``systematic error''.
}.
We consider these oscillations (along with the fact that the spectra invariably peak at the largest scale rather than the local scale height) as artefacts of the shell model, rather than as relevant to full-fledged convection.

On a more speculative note, we see similarities between the behaviour of our shell model and that expected from the entropy rain picture.
If convection in the Sun were indeed driven by downflowing plumes or thermals, the effect of low \Pr{} could be understood in two ways: 
\begin{enuminline}
	\item a low viscosity leads to less dissipation of the downflows' momentum, allowing them to persist longer
	\item a high thermal diffusivity leads to downflows losing their entropy signature very quickly, leading to a suppression of buoyant braking.
\end{enuminline}
These seem to be consistent with the suggestion of \citet{kapyla2021Pr} that as \Pr{} is lowered, the kinetic energy of the flows increases. 

Recently, \citet{Fuentes2022} have found that at low \Pr{}, convective layers in simulated gas giant planets start merging.
The reason for their observation is unclear, but it may be related to the mechanism at play in the Sun's convection zone.

We also note that \citet{vlaykov2022}, in 2D simulations of a solar-like convective region, have found that the near-surface layers affect even the bottom of the convection zone.
The apparent disagreement between their results and those of \citet{hotta2019weakSurface} can be explained as follows:
\citet{hotta2019weakSurface} used a slope-limited diffusion scheme which dissipates both the momentum and the entropy in a similar fashion.
Their results thus represent convection at $\Pr \orderOf 1$.
On the other hand, \citet{vlaykov2022} performed an implicit LES of the momentum equation, while retaining an explicit thermal diffusivity.
Their results then seem to represent convection at low \Pr{} \citep{bricteux2012}.
This supports our finding that a low \Pr{} enhances the effect of a strongly stratified unstable layer on the convective velocities in deeper layers.

\section{Conclusions}
\label{section: conclusions}

We find that a low value of \Pr{}, along with a thermal-diffusion-induced coupling between different depths, is sufficient to cause suppression of convective spectra as one goes deeper into a convecting domain underneath a strongly unstable surface layer. 
This suggests that low \Pr{} is at least as important as magnetic fields and rotation for a complete explanation of the convective conundrum.

Given that \citet{Featherstone2016HighRa} report suppression of large scales in their simulations as \Ra{} increases, we also note the possibility that large \Ra{} and low \Pr{} are degenerate, and that some combination of the two serves as a better control parameter.
In light of our findings, the effects of low \Pr{} on the stability and dynamics of convective structures (such as the plumes invoked in the entropy rain picture) should be studied.
These will be taken up elsewhere.

\backsection[Acknowledgments]{
We thank Axel Brandenburg and Petri K\"apyl\"a for valuable discussions on solar convection.
We thank Alexandra Elbakyan for facilitating access to scientific literature during the recent pandemic.
We acknowledge use of the Pegasus computing cluster at IUCAA.
}

\backsection[Funding]{
This research received no specific grant from any funding agency, commercial or not-for-profit sectors.
}

\backsection[Software]{
Scipy \citep{scipy2020}, Numpy \citep{numpy2020}, and Matplotlib \citep{matplotlib2007}.
}

\backsection[Declaration of interests]{
The authors report no conflict of interest.
}

\backsection[Author ORCID]{
KG, \url{https://orcid.org/0000-0003-2620-790X}; 
NS, \url{https://orcid.org/0000-0001-6097-688X}
}

\backsection[Author contributions]{
KG and NS conceptualized the research.
KG constructed the model and performed simulations.
KG and NS interpreted the results and wrote the paper.
}

\appendix
\section{Dimensionless numbers in the shell model}
\label{appendix: Pr Ra in shell model}
Setting $C_P = \kappa = 1$, we write
\begin{equation}
	\Pr = \nu
\end{equation}
The Rayleigh number may be estimated as
\begin{equation}
	\Ra = \frac{ g \left( \sh{s}{0}{M} - \sh{s}{0}{1} \right) \left( M \shdZ \right)^3 }{ \nu }
	\label{CO2.StratSh: eq: Rayleigh number estimate}
\end{equation}
assuming that $\sh{s}{0}{M} > \sh{s}{0}{1}$ (i.e. that the domain is unstably stratified).
Recalling that $\Ra = \Gr \Pr$ and $\Gr = \R^2$ \citep[eq.~17]{SchumacherSreenivasanSolConv2020}, we can estimate the Reynolds number as 
\begin{equation}
	\R = \sqrt{ \frac{\Ra}{\Pr} }
	\label{CO2.StratSh: eq: Reynolds number estimate}
\end{equation}

\section{Varying the Prandtl number}
\label{appendix: vary Pr keeping Ra const}
We vary $\nu$ to change \Pr{} (see appendix \ref{appendix: Pr Ra in shell model}).
However, this also changes \Ra{}.
To vary \Pr{} while keeping \Ra{} unchanged, we simultaneously modify
\begin{align}
	\Pr &\to K\Pr
	\\
	g &\to K g
\end{align}
where $K$ is some positive constant.

Empirically, we find that with the above modifications, the turnover timescale roughly scales as
\begin{equation}
	\tau \to K^{-1/2} \tau
\end{equation}
which is consistent with estimating it from the free-fall velocity:
\begin{equation}
	\tau 
	\orderOf \left( \frac{g \grad s_0}{C_P} \right)^{-1/2}
	\label{CO2.StratSh.Ma0: estimate convective turnover time}
\end{equation}

\bibliographystyle{jfm}
\bibliography{shell2022.bib}

\end{document}